\documentclass{iau}

\usepackage{amsmath}
\usepackage{graphicx}
\usepackage{multirow}

\begin{document}

\lefttitle{ALMA and EVLA study of I19520}
\righttitle{Proceedings of the International Astronomical Union}

\jnlPage{1}{7}
\jnlDoiYr{2026}
\doival{10.1017/xxxxx}

\aopheadtitle{Proceedings IAU Symposium}
\editors{A. Wofford,  N. St-Louis, M. Garcia \&  S. Simón-Díaz, eds}
\title{Unveiling the First O-Type `Bloated Star’ Candidate through ALMA and EVLA Observations}
\author{Rakesh Pandey$^{1}$\thanks{pandey.rakesh405@gmail.com}, Aina Palau$^{1}$, Alvaro S\'anchez-Monge$^{2,3}$, Raghvendra Sahai$^{4}$, Rolf Kuiper$^{5}$, Luis Felipe Rodr\'iguez$^{1}$, Carmen S\'anchez Contreras$^{6}$, Saurabh Sharma$^{7}$}
\affiliation{$^{1}$ Instituto de Radioastronom\'ia y Astrof\'isica, UNAM, Morelia, Mexico; 
$^{2}$ Institut de Ci\`encies de l'Espai (ICE--CSIC), Bellaterra, Spain; $^{3}$ Institut d'Estudis Espacials de Catalunya (IEEC), Barcelona, Spain; $^{4}$ Jet Propulsion Laboratory, California Institute of Technology, Pasadena, USA; $^{5}$ Faculty of Physics, University of Duisburg--Essen, Duisburg, Germany; $^{6}$ Centro de Astrobiolog\'ia (CAB), CSIC--INTA, Madrid, Spain; $^{7}$ Aryabhatta Research Institute of Observational Sciences (ARIES), Nainital, India}
\begin{abstract}
We investigate the circumstellar environment of the O-type bloated star candidate IRAS 19520+2759 (I19520) using high-resolution observations from the Atacama Large Millimeter/submillimeter Array (ALMA) and the Expanded Very Large Array (EVLA). Radio continuum emission traced by the EVLA (C, K, and Q bands) exhibits a spectral index of 0.5, consistent with a thermal jet. ALMA 1.3 mm continuum map reveals a compact source coincident with the optical counterpart of I19520, likely tracing the dense core hosting the central massive young stellar object. A prominent molecular outflow in the east–west direction, along with a possible secondary outflow oriented northeast–southwest, is identified in the $^{13}$CO emission. A hot molecular core and a Keplerian disk are detected in several SO$_2$ transitions. Assuming an edge-on disk geometry, the dynamical mass of the central object is estimated to be in the range of 10--15 M$\odot$.
\end{abstract}
\maketitle
\section{Introduction}
Bloated stars are high-mass young stellar objects (MYSO) in a crucial evolutionary phase of active accretion– when the stellar surface is expanding because of the accretion of matter on to the star \citep{hos10}. These stars, unlike typical O-type stars on the main-sequence, have a relatively low effective temperature (consistent with a B-spectral type) and low ionizing flux, and thus are highly deficient in radio emission. This theoretically predicted “bloated” or “swollen” stage in massive star formation remains poorly constrained observationally, with only a few candidates identified so far. Among them, IRAS 19520+2759 (I19520) is the most luminous (L $>$ 10$^5$ L$_\odot$) and represents the first O-type bloated star candidate \citep{aina13}.
I19520 (RA (J2000) =19h54m05.9s, Dec(J2000)=+28d7m41s) is  a MYSO, located at 9 kpc, and detected in optical and Near-infrared (NIR) wavelengths \citep{car13}. The source is classified as “Bloated star candidate” based on its high luminosity and relatively low effective temperature estimated from its optical spectrum and lack of centimeter radio emission \citep{aina13}. In our recent investigation, we performed a variability study on I19520 using optical and infrared time series data and found that it shows periodic variability consistent with the pulsational variability predicted for bloated sources \citep{pandey25}. 
Interferometric (OVRO) millimeter observations revealed a compact dusty core coincident with the optical source, a powerful collimated CO outflow, and an elongated C$^{18}$O structure perpendicular to it \citep{aina13}. However, these observations lacked the angular resolution necessary to probe the inner structure of the system. To overcome these limitations, we conducted high angular resolution observations with the Atacama Large Millimeter/submillimeter Array (ALMA; beam size $\sim 0.18^{\prime\prime} \times 0.13^{\prime\prime}$) and the Expanded Very Large Array (EVLA; beam size at Q band $\sim 0.41^{\prime\prime} \times 0.37^{\prime\prime}$). Our aim is to present the first detailed study of the physical and chemical properties of the dense circumstellar environment of an O-type bloated star candidate. 
\section{Results and Discussion} \label{sec1}
\begin{figure*}
    \includegraphics[scale=.45]{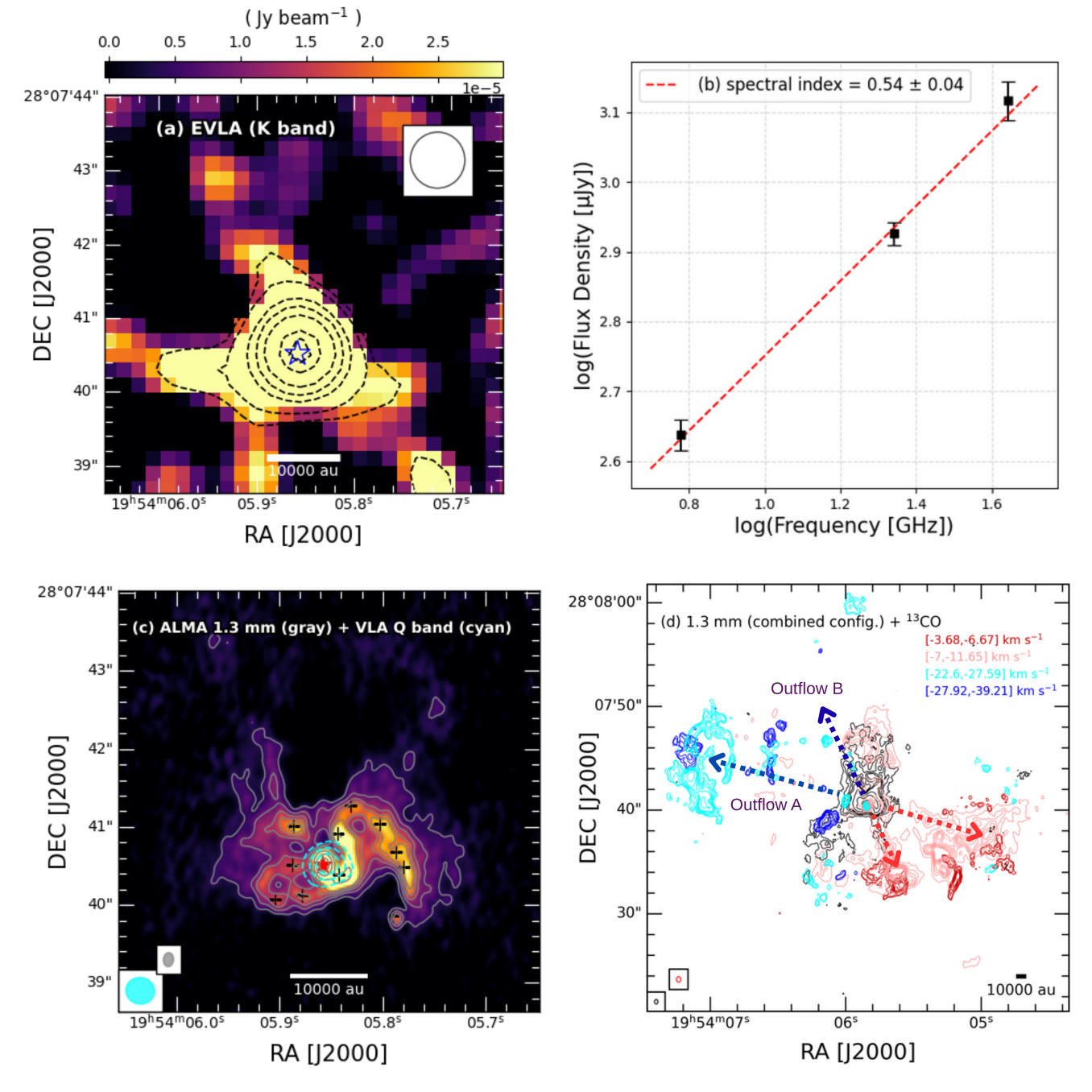}
    \caption{ (a) EVLA K band image showing the radio continuum emission associated with I19520. Contours are drawn at 3, 5, 13, 20, 30, 50, and 60 $\times$(rms = 9$\times$10$^{-6}$ Jy beam$^{-1}$). (b) Flux density versus frequency plot derived from the EVLA C, K, and Q bands. The slope of the fitted line yields the spectral index of the radio emission. (c) ALMA 1.3 mm continuum image with major continuum peaks are marked with plus symbols. Contours are drawn at 3, 5, 7, 9, 15, 20, 25, 32, 50, and 100$\times$rms, where rms $\sim$ 3.4 mJy beam$^{-1}$. (d) Blue- and redshifted $^{13}$CO emission (see Section \ref{sec1}) shown in contours, overlaid with 1.3 mm continuum contours (black). Arrows indicate the directions of the molecular outflows identified in this work. The star symbols in panels (a) and (c) mark the position of the optical counterpart of I19520.}
\label{fig1}
\end{figure*}

\begin{figure*}
    \includegraphics[scale=.5]{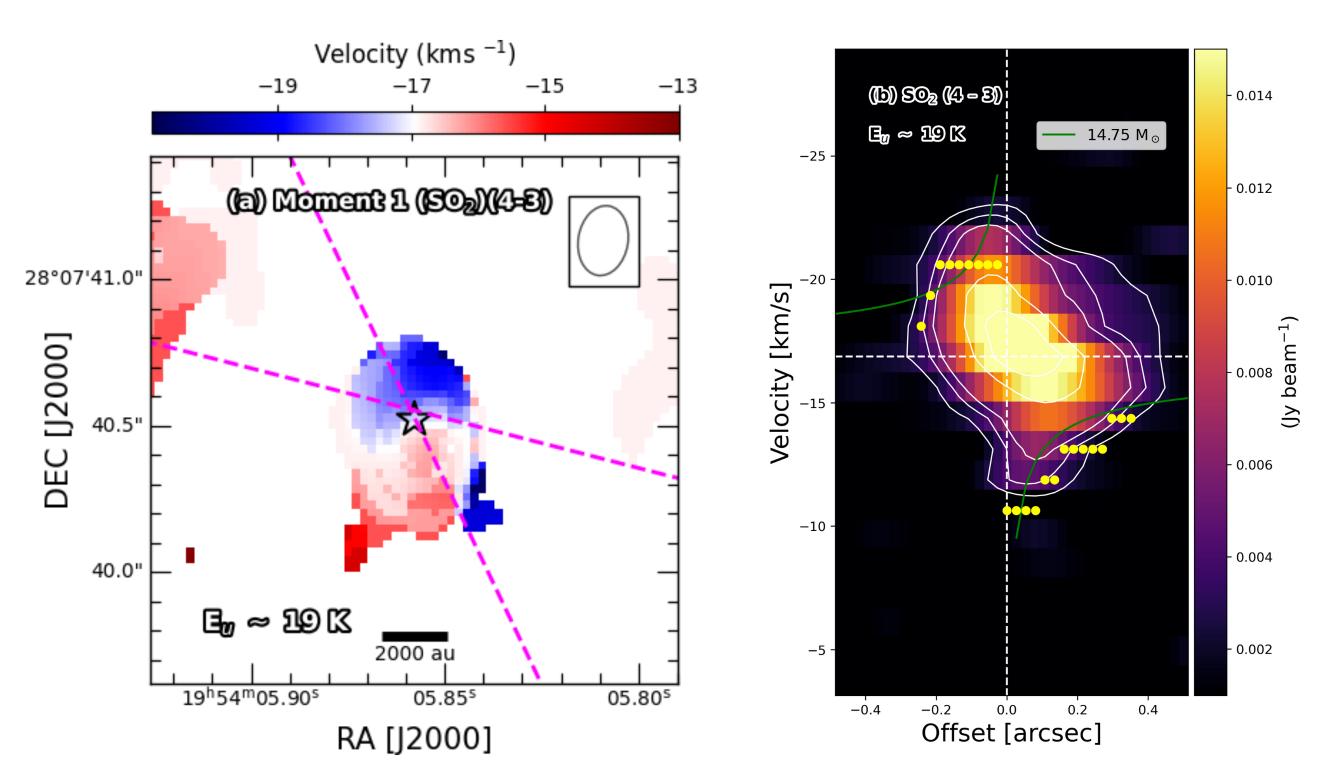}
    \caption{(a) Moment-1 map of the SO$_2$ (4–3) transition, with lines indicating the directions of the plausible outflows (see Section \ref{sec1}); the star marks the position of the continuum peak MM1. (b) PV plot for  SO$_2$ (4–3) transition, extracted along the velocity gradient in Moment-1 map, the contours (white) are drawn at 3, 5, 7, 15, 20$\times$(rms=8.6$\times$$10^{-4}$ Jy beam$^{-1}$). The points mark the maximum velocity for different radial offsets shown along with the best fitted Keplerian rotation model (green).}
    \label{fig2}
  \end{figure*}

\textbf{Continuum emission and molecular outflows}:
Figure \ref{fig1} (a) shows the EVLA K band image  (beam size $\sim 0.76^{\prime\prime} \times 0.74^{\prime\prime}$) tracing the radio continuum emission associated with I19520. The emission exhibits an extended morphology, elongated along the east–west and northeast–southwest directions, and peaks at the position of the optical counterpart of I19520. The spectral index, derived from the EVLA C, K, and Q band fluxes, is 0.54 ± 0.04, consistent with a thermal radio jet.
We show the ALMA Band 6 (1.3 mm) continuum in Figure \ref{fig1} (c). The 1.3 mm continuum map reveals a complex, asymmetric morphology with a bright, compact source at the center (MM1). The peak of this compact emission coincides closely with the optical counterpart (similar to the radio continuum emission), likely tracing the core that harbors the MYSO I19520. In this study, we use $^{13}$CO as a tracer of the molecular outflow. Figure \ref{fig1} (d) presents the $^{13}$CO emission integrated over velocity ranges excluding the systemic velocity interval centered at -13.6 km s$^{-1}$. The blueshifted lobe spans –35.0 to –22.06 km s$^{-1}$, and the redshifted lobe –11.68 to –6.0 km s$^{-1}$. The high-velocity $^{13}$CO emission exhibits a clear bipolar morphology, with the blue- and redshifted lobes extending in opposite directions from MM1, oriented roughly east–west ($\mathrm{PA} \approx 78^\circ$, hereafter Outflow A). Additionally, another high-velocity $^{13}$CO bipolar feature is observed along the northeast–southwest direction ($\mathrm{PA} \approx 28^\circ$, hereafter Outflow B; see Figure \ref{fig1} (d)), although its blueshifted component is relatively weak.
\textbf{Hot core and a Keplerian disk}: 
Figure \ref{fig2} (a) shows the intensity-weighted velocity (Moment-1) map of the ALMA SO$_{2}$ (4–3) emission. The compact emission centered on the continuum source MM1, likely traces the hot-core chemistry, as sulphur-bearing species such as SO$_{2}$ are commonly enhanced in hot molecular cores \citep[e.g.,][]{ham24}. 

A clear north-south velocity gradient, roughly perpendicular to outflow A, suggests rotational motion within the hot core. Such rotation is often attributed to toroids or circumstellar disks exhibiting Keplerian rotation \citep[e.g.,][]{bel17}.
To investigate the kinematics in more detail, we constructed a position–velocity (PV) diagram along the observed gradient (Figure \ref{fig2} (b)) and fitted the velocity distribution with Keplerian rotation curves following \citet{sef16}. Performing the same analysis for higher SO$2$ transitions, we derive a dynamical mass in the range of 10–15 M$\odot$, consistent with the findings of \citet{pandey25} for the disk-accretion case.

\end{document}